\documentclass[twocolumn,secnumarabic,amssymb, nobibnotes, aps, prl,floatfix,superscriptaddress]{revtex4-1}

\usepackage{graphicx} 
\usepackage{hyperref}

\begin{document}

\title{Demonstration of Single Barium Ion Sensitivity for Neutrinoless Double Beta Decay using Single Molecule Fluorescence Imaging}

\collaboration{The NEXT Collaboration}

\author{A.D.McDonald}
\thanks{Corresponding author: \url{austin.mcdonald@uta.edu}}
\affiliation{Department of Physics, University of Texas at Arlington,
Arlington, Texas 76019, USA}
\author{B.J.P. Jones}
\thanks{Corresponding author: \url{ben.jones@uta.edu}}
\affiliation{Department of Physics, University of Texas at Arlington,
Arlington, Texas 76019, USA}
\author{D.R.Nygren}
\thanks{NEXT Co-spokesperson.}
\affiliation{Department of Physics, University of Texas at Arlington,
Arlington, Texas 76019, USA}


\author{C.~Adams}
\affiliation{Department of Physics, Harvard University, Cambridge, MA 02138, USA}
\author{V.~\'Alvarez}
\affiliation{Instituto de F\'isica Corpuscular (IFIC), CSIC \& Universitat de Val\`encia, 
Calle Catedr\'atico Jos\'e Beltr\'an, 2, 46980 Paterna, Valencia, Spain}
\author{C.D.R.~Azevedo}
\affiliation{Institute of Nanostructures, Nanomodelling and Nanofabrication (i3N), Universidade de Aveiro, 
Campus de Santiago, 3810-193 Aveiro, Portugal}
\author{J.M.~Benlloch-Rodr\'{i}guez}
\affiliation{Instituto de F\'isica Corpuscular (IFIC), CSIC \& Universitat de Val\`encia, 
Calle Catedr\'atico Jos\'e Beltr\'an, 2, 46980 Paterna, Valencia, Spain}
\author{F.I.G.M.~Borges}
\affiliation{LIP, Department of Physics, University of Coimbra, 
P-3004 516 Coimbra, Portugal}
\author{A.~Botas}
\affiliation{Instituto de F\'isica Corpuscular (IFIC), CSIC \& Universitat de Val\`encia, 
Calle Catedr\'atico Jos\'e Beltr\'an, 2, 46980 Paterna, Valencia, Spain}
\author{S.~C\'arcel}
\affiliation{Instituto de F\'isica Corpuscular (IFIC), CSIC \& Universitat de Val\`encia, 
Calle Catedr\'atico Jos\'e Beltr\'an, 2, 46980 Paterna, Valencia, Spain}
\author{J.V.~Carri\'on}
\affiliation{Instituto de F\'isica Corpuscular (IFIC), CSIC \& Universitat de Val\`encia, 
Calle Catedr\'atico Jos\'e Beltr\'an, 2, 46980 Paterna, Valencia, Spain}
\author{S.~Cebri\'an}
\affiliation{Laboratorio de F\'isica Nuclear y Astropart\'iculas, Universidad de Zaragoza, 
Calle Pedro Cerbuna, 12, 50009 Zaragoza, Spain}
\author{C.A.N.~Conde}
\affiliation{LIP, Department of Physics, University of Coimbra, 
P-3004 516 Coimbra, Portugal}
\author{J.~D\'iaz}
\affiliation{Instituto de F\'isica Corpuscular (IFIC), CSIC \& Universitat de Val\`encia, 
Calle Catedr\'atico Jos\'e Beltr\'an, 2, 46980 Paterna, Valencia, Spain}
\author{M.~Diesburg}
\affiliation{Fermi National Accelerator Laboratory, 
Batavia, Illinois 60510, USA}
\author{J.~Escada}
\affiliation{LIP, Department of Physics, University of Coimbra, 
P-3004 516 Coimbra, Portugal}
\author{R.~Esteve}
\affiliation{Instituto de Instrumentaci\'on para Imagen Molecular (I3M), Centro Mixto CSIC -â€” Universitat Polit\`ecnica de Val\`encia, 
Camino de Vera s/n, 46022 Valencia, Spain}
\author{R.~Felkai}
\affiliation{Instituto de F\'isica Corpuscular (IFIC), CSIC \& Universitat de Val\`encia, 
Calle Catedr\'atico Jos\'e Beltr\'an, 2, 46980 Paterna, Valencia, Spain}
\author{L.M.P.~Fernandes}
\affiliation{LIBPhys, Physics Department, University of Coimbra, 
Rua Larga, 3004-516 Coimbra, Portugal}
\author{P.~Ferrario}
\affiliation{Instituto de F\'isica Corpuscular (IFIC), CSIC \& Universitat de Val\`encia, 
Calle Catedr\'atico Jos\'e Beltr\'an, 2, 46980 Paterna, Valencia, Spain}
\author{A.L.~Ferreira}
\affiliation{Institute of Nanostructures, Nanomodelling and Nanofabrication (i3N), Universidade de Aveiro, 
Campus de Santiago, 3810-193 Aveiro, Portugal}
\author{E.D.C.~Freitas}
\affiliation{LIBPhys, Physics Department, University of Coimbra, 
Rua Larga, 3004-516 Coimbra, Portugal}
\author{A.~Goldschmidt}
\affiliation{Lawrence Berkeley National Laboratory (LBNL), 
 Berkeley, California 94720, USA}
\author{J.J.~G\'omez-Cadenas}
\thanks{NEXT Co-spokesperson.}
\affiliation{Instituto de F\'isica Corpuscular (IFIC), CSIC \& Universitat de Val\`encia, 
Calle Catedr\'atico Jos\'e Beltr\'an, 2, 46980 Paterna, Valencia, Spain}
\author{D.~Gonz\'alez-D\'iaz}
\affiliation{Instituto Gallego de F\'isica de Altas Energ\'ias, Univ.\ de Santiago de Compostela, 
Campus sur, R\'ua Xos\'e Mar\'ia Su\'arez N\'u\~nez, s/n, 15782 Santiago de Compostela, Spain}
\author{R.M.~Guti\'errez}
\affiliation{Centro de Investigaci\'on en Ciencias B\'asicas y Aplicadas, Universidad Antonio Nari\~no, 
Sede Circunvalar, Carretera 3 Este No.\ 47 A-15, Bogot\'a, Colombia}
\author{R.~Guenette}
\affiliation{Department of Physics, Harvard University, Cambridge, MA 02138, USA}
\author{K.~Hafidi}
\affiliation{Argonne National Laboratory, Argonne IL 60439, USA}
\author{J.~Hauptman}
\affiliation{Department of Physics and Astronomy, Iowa State University, 
 Ames, Iowa 50011-3160, USA}
\author{C.A.O.~Henriques}
\affiliation{LIBPhys, Physics Department, University of Coimbra, 
Rua Larga, 3004-516 Coimbra, Portugal}
\author{A.I.~Hernandez}
\affiliation{Centro de Investigaci\'on en Ciencias B\'asicas y Aplicadas, Universidad Antonio Nari\~no,  
Sede Circunvalar, Carretera 3 Este No.\ 47 A-15, Bogot\'a, Colombia}
\author{J.A.~Hernando~Morata}
\affiliation{Instituto Gallego de F\'isica de Altas Energ\'ias, Univ.\ de Santiago de Compostela, 
Campus sur, R\'ua Xos\'e Mar\'ia Su\'arez N\'u\~nez, s/n, 15782 Santiago de Compostela, Spain}
\author{V.~Herrero}
\affiliation{Instituto de Instrumentaci\'on para Imagen Molecular (I3M), Centro Mixto CSIC -â€” Universitat Polit\`ecnica de Val\`encia, 
Camino de Vera s/n, 46022 Valencia, Spain}
\author{S.~Johnston}
\affiliation{Argonne National Laboratory, Argonne IL 60439, USA}
\author{L.~Labarga}
\affiliation{Departamento de F\'isica Te\'orica, Universidad Aut\'onoma de Madrid, 
Campus de Cantoblanco, 28049 Madrid, Spain}
\author{A.~Laing}
\affiliation{Instituto de F\'isica Corpuscular (IFIC), CSIC \& Universitat de Val\`encia, 
Calle Catedr\'atico Jos\'e Beltr\'an, 2, 46980 Paterna, Valencia, Spain}
\author{P.~Lebrun}
\affiliation{Fermi National Accelerator Laboratory,  
Batavia, Illinois 60510, USA}
\author{I.~Liubarsky}
\affiliation{Instituto de F\'isica Corpuscular (IFIC), CSIC \& Universitat de Val\`encia, 
Calle Catedr\'atico Jos\'e Beltr\'an, 2, 46980 Paterna, Valencia, Spain}
\author{N.~L\'opez-March}
\affiliation{Department of Physics, University of Texas at Arlington, 
Arlington, Texas 76019, USA}
\affiliation{Instituto de F\'isica Corpuscular (IFIC), CSIC \& Universitat de Val\`encia, 
Calle Catedr\'atico Jos\'e Beltr\'an, 2, 46980 Paterna, Valencia, Spain}
\author{M.~Losada}
\affiliation{Centro de Investigaci\'on en Ciencias B\'asicas y Aplicadas, Universidad Antonio Nari\~no, 
Sede Circunvalar, Carretera 3 Este No.\ 47 A-15, Bogot\'a, Colombia}
\author{J.~Mart\'in-Albo}
\affiliation{Department of Physics, Harvard University, Cambridge, MA 02138, USA}
\author{G.~Mart\'inez-Lema}
\affiliation{Instituto Gallego de F\'isica de Altas Energ\'ias, Univ.\ de Santiago de Compostela, 
Campus sur, R\'ua Xos\'e Mar\'ia Su\'arez N\'u\~nez, s/n, 15782 Santiago de Compostela, Spain}
\author{A.~Mart\'inez}
\affiliation{Instituto de F\'isica Corpuscular (IFIC), CSIC \& Universitat de Val\`encia, 
Calle Catedr\'atico Jos\'e Beltr\'an, 2, 46980 Paterna, Valencia, Spain}
\author{F.~Monrabal}
\affiliation{Department of Physics, University of Texas at Arlington,
Arlington, Texas 76019, USA}
\author{C.M.B.~Monteiro}
\affiliation{LIBPhys, Physics Department, University of Coimbra,
Rua Larga, 3004-516 Coimbra, Portugal}
\author{F.J.~Mora}
\affiliation{Instituto de Instrumentaci\'on para Imagen Molecular (I3M), Centro Mixto CSIC -â€” Universitat Polit\`ecnica de Val\`encia,
Camino de Vera s/n, 46022 Valencia, Spain}
\author{L.M.~Moutinho}
\affiliation{Institute of Nanostructures, Nanomodelling and Nanofabrication (i3N), Universidade de Aveiro,
Campus de Santiago, 3810-193 Aveiro, Portugal}
\author{J.~Mu\~noz Vidal}
\affiliation{Instituto de F\'isica Corpuscular (IFIC), CSIC \& Universitat de Val\`encia,
Calle Catedr\'atico Jos\'e Beltr\'an, 2, 46980 Paterna, Valencia, Spain}
\author{M.~Musti}
\affiliation{Instituto de F\'isica Corpuscular (IFIC), CSIC \& Universitat de Val\`encia,
Calle Catedr\'atico Jos\'e Beltr\'an, 2, 46980 Paterna, Valencia, Spain}
\author{M.~Nebot-Guinot}
\affiliation{Instituto de F\'isica Corpuscular (IFIC), CSIC \& Universitat de Val\`encia,
Calle Catedr\'atico Jos\'e Beltr\'an, 2, 46980 Paterna, Valencia, Spain}
\author{P.~Novella}
\affiliation{Instituto de F\'isica Corpuscular (IFIC), CSIC \& Universitat de Val\`encia,
Calle Catedr\'atico Jos\'e Beltr\'an, 2, 46980 Paterna, Valencia, Spain}
\author{B.~Palmeiro}
\affiliation{Instituto de F\'isica Corpuscular (IFIC), CSIC \& Universitat de Val\`encia,
Calle Catedr\'atico Jos\'e Beltr\'an, 2, 46980 Paterna, Valencia, Spain}
\author{A.~Para}
\affiliation{Fermi National Accelerator Laboratory, 
Batavia, Illinois 60510, USA}
\author{J.~P\'{e}rez}
\affiliation{Instituto de F\'isica Corpuscular (IFIC), CSIC \& Universitat de Val\`encia,
Calle Catedr\'atico Jos\'e Beltr\'an, 2, 46980 Paterna, Valencia, Spain}
\author{M.~Querol}
\affiliation{Instituto de Instrumentaci\'on para Imagen Molecular (I3M), Centro Mixto CSIC -â€” Universitat Polit\`ecnica de Val\`encia,
Camino de Vera s/n, 46022 Valencia, Spain}
\author{J. Repond}
\affiliation{Argonne National Laboratory, Argonne IL 60439, USA}
\author{J.~Renner}
\affiliation{Instituto de F\'isica Corpuscular (IFIC), CSIC \& Universitat de Val\`encia,
Calle Catedr\'atico Jos\'e Beltr\'an, 2, 46980 Paterna, Valencia, Spain}
\author{S. Riordan}
\affiliation{Argonne National Laboratory, Argonne IL 60439, USA}
\author{L.~Ripoll}
\affiliation{Escola Polit\`ecnica Superior, Universitat de Girona,
Av.~Montilivi, s/n, 17071 Girona, Spain}
\author{J.~Rodr\'iguez}
\affiliation{Instituto de F\'isica Corpuscular (IFIC), CSIC \& Universitat de Val\`encia,
Calle Catedr\'atico Jos\'e Beltr\'an, 2, 46980 Paterna, Valencia, Spain}
\author{L.~Rogers}
\affiliation{Department of Physics, University of Texas at Arlington,
Arlington, Texas 76019, USA}
\author{F.P.~Santos}
\affiliation{LIP, Department of Physics, University of Coimbra,
P-3004 516 Coimbra, Portugal}
\author{J.M.F.~dos~Santos}
\affiliation{LIBPhys, Physics Department, University of Coimbra,
Rua Larga, 3004-516 Coimbra, Portugal}
\author{A.~Sim\'on}
\affiliation{Instituto de F\'isica Corpuscular (IFIC), CSIC \& Universitat de Val\`encia,
Calle Catedr\'atico Jos\'e Beltr\'an, 2, 46980 Paterna, Valencia, Spain}
\author{C.~Sofka}
\thanks{Now at University of Texas at Austin, USA.}
\author{M.~Sorel}
\affiliation{Instituto de F\'isica Corpuscular (IFIC), CSIC \& Universitat de Val\`encia,
Calle Catedr\'atico Jos\'e Beltr\'an, 2, 46980 Paterna, Valencia, Spain}
\author{T.~Stiegler}
\affiliation{Department of Physics and Astronomy, Texas A\&M University,
College Station, Texas 77843-4242, USA}
\author{J.F.~Toledo}
\affiliation{Instituto de Instrumentaci\'on para Imagen Molecular (I3M), Centro Mixto CSIC -â€” Universitat Polit\`ecnica de Val\`encia,
Camino de Vera s/n, 46022 Valencia, Spain}
\author{J.~Torrent}
\affiliation{Instituto de F\'isica Corpuscular (IFIC), CSIC \& Universitat de Val\`encia,
Calle Catedr\'atico Jos\'e Beltr\'an, 2, 46980 Paterna, Valencia, Spain}
\author{Z.~Tsamalaidze}
\affiliation{Joint Institute for Nuclear Research (JINR),
Joliot-Curie 6, 141980 Dubna, Russia}
\author{J.F.C.A.~Veloso}
\affiliation{Institute of Nanostructures, Nanomodelling and Nanofabrication (i3N), Universidade de Aveiro,
Campus de Santiago, 3810-193 Aveiro, Portugal}
\author{R.~Webb}
\affiliation{Department of Physics and Astronomy, Texas A\&M University,
College Station, Texas 77843-4242, USA}
\author{J.T.~White}
\thanks{Deceased.}
\affiliation{Department of Physics and Astronomy, Texas A\&M University,
College Station, Texas 77843-4242, USA}
\author{N.~Yahlali}
\affiliation{Instituto de F\'isica Corpuscular (IFIC), CSIC \& Universitat de Val\`encia,
Calle Catedr\'atico Jos\'e Beltr\'an, 2, 46980 Paterna, Valencia, Spain}
%

\date{November 11, 2017}

 \begin{abstract}
A new method to tag the barium daughter in the double beta decay of $^{136}$Xe is reported.  Using the technique of single molecule fluorescent imaging (SMFI), individual barium dication (Ba$^{++}$) resolution at a transparent scanning surface has been demonstrated.  A single-step photo-bleach confirms the single ion interpretation.  Individual ions are localized with super-resolution ($\sim$2~nm), and detected with a statistical significance of 12.9~$\sigma$ over backgrounds.  This lays the foundation for a new and potentially background-free neutrinoless double beta decay technology, based on SMFI coupled to high pressure xenon gas time projection chambers.
 \end{abstract}
 
\maketitle
\flushbottom

\section{Introduction}

The nature of neutrino mass is one of the fundamental open questions in nuclear and particle physics.  If neutrinos are Majorana particles, their tiny mass may be evidence for high energy-scale physics via the seesaw mechanism \cite{Chang:1985en,minkowski1977mu,gell1979ramond,yanagida1979proceedings,mohapatra1981neutrino}, and lend support for a compelling theoretical explanation of the matter-antimatter imbalance in the universe (leptogenesis) \cite{Fukugita:1986hr}.  The most sensitive known method to establish the Majorana nature of the neutrino experimentally is direct observation of neutrinoless double beta decay ($0\nu\beta\beta$) \cite{Ostrovskiy:2016uyx,DellOro:2016tmg, GomezCadenas:2010gs,gomez2011search}, a radioactive process that can occur if and only if the neutrino is a Majorana fermion.  The mass scale implied by direct limits \cite{Aseev:2011dq}, cosmology \cite{Ade:2015xua} and neutrino oscillations \cite{Gonzalez-Garcia:2015qrr} dictates that the rate of $0\nu\beta\beta$, assuming the standard mechanism, will be very low: the next generation of experiments must probe $0\nu\beta\beta$ lifetimes of $\geq 10^{27}$ years. Observation of such a rare decay requires ton-scale detectors with near-perfect background rejection capabilities. 

One candidate $0\nu\beta\beta$ isotope which has been a focus of much attention is $^{136}$Xe.  As a noble element, xenon can be used in time projection chamber (TPC) detectors, enabling fully active, monolithic $0\nu\beta\beta$ searches in both gas \cite{Alvarez:2012flf} and liquid \cite{Auger:2012ar} phases.  Background measurements in present-generation detectors at the 10-100~kg scale, however, suggest that all proposed ton-scale experiments will remain background limited in their sensitivity.  Improved technologies with enhanced background rejection capabilities will be required in order to make further progress as well as enhance confidence for a discovery claim.

It has long been recognized that the detection of single barium ions emanating from the decay of $^{136}$Xe \cite{Moe:1991ik}, when combined with a Gaussian energy resolution better than 2\% FWHM ($\sigma/E\sim0.85\%)$ to reject the two-neutrino-mode background, could enable a background-free $0\nu\beta\beta$ search. This is because no conventional radioactive process can produce a barium ion in bulk xenon.  So-called ``barium tagging'' has been a subject of R\&D for at least 20 years \cite{Danilov:2000pp,Mong:2014iya,Rollin:2011gla,Brunner:2014sfa,Flatt:2007aa,Sinclair:2011zz}, but at the time of writing, a convincing method of barium ion extraction and identification remains elusive.

Single molecule fluorescence imaging (SMFI) is a technique invented by physicists and developed by biochemists that enables single-molecule sensitive, super-resolution microscopy. Among the applications of SMFI are the sensing of individual ions \cite{Lu2007}, demonstrated in various environments, including inside living cells \cite{stuurman2006imaging}. A fluor is employed that is non-fluorescent in isolation, but becomes fluorescent upon chelation with a suitable ion.  The molecule typically comprises of a dye bonded to a receptor that traps the ion in a cage-like structure. The electrostatic forces exhibited by the ion on the dye modify its energy levels to enable molecular fluorescence.   Detection is assisted by the inherent stokes shift of the dye, allowing separation of emission and excitation light via dichroic filters.  Localized light emission from single molecules can be spatially resolved using electron multiplying CCD (EM-CCD) cameras, allowing rejection of backgrounds from scattering and low-level fluorescence of unchelated molecules, which are diffuse.

The NEXT collaboration \cite{Alvarez:2012flf} is pursuing a program of R\&D to employ SMFI techniques to detect the barium daughter ion in high pressure xenon gas (HPGXe). In HPGXe, energy resolutions extrapolating to better than 1\%~FWHM at Q$_{\beta\beta}$ have been achieved \cite{Alvarez:2012yxw}, sufficient to efficiently reject the two-neutrino-mode background.  Barium resulting from double beta decay is initially highly ionized due to the disruptive departure of the two energetic electrons from the nucleus \cite{PhysRev.107.1646}. Rapid capture of electrons from neutral xenon is expected to reduce this charge state to Ba$^{++}$, which may then be further neutralized through electron-ion recombination.  Unlike in liquid xenon, where recombination is frequent and  the barium daughters are distributed across charge states~\cite{PhysRevC.92.045504}, recombination in the gas phase is minimal \cite{1997NIMPA.396..360B}, and thus Ba$^{++}$ is the expected outcome.  Furthermore, since the ion energy in high pressure gas is thermal, and charge exchange with xenon is highly energetically disfavored, the Ba$^{++}$ state is expected to persist through drift to the anode plane.  For this reason, and because barium and calcium are congeners, dyes which have been developed for Ca$^{++}$ sensitivity for biochemistry applications provide a promising path toward barium tagging in HPGXe.  In ref. \cite{Jones:2016qiq} we explored the properties of two such dyes, Fluo-3 and Fluo-4.  In the presence of Ba$^{++}$, excitation at 488~nm yielded strong emission peaking around 525~nm, demonstrating the potential of these dyes to serve as barium tagging agents.

\begin{figure}[b]
\label{Lego}
\includegraphics[width=0.8\columnwidth]{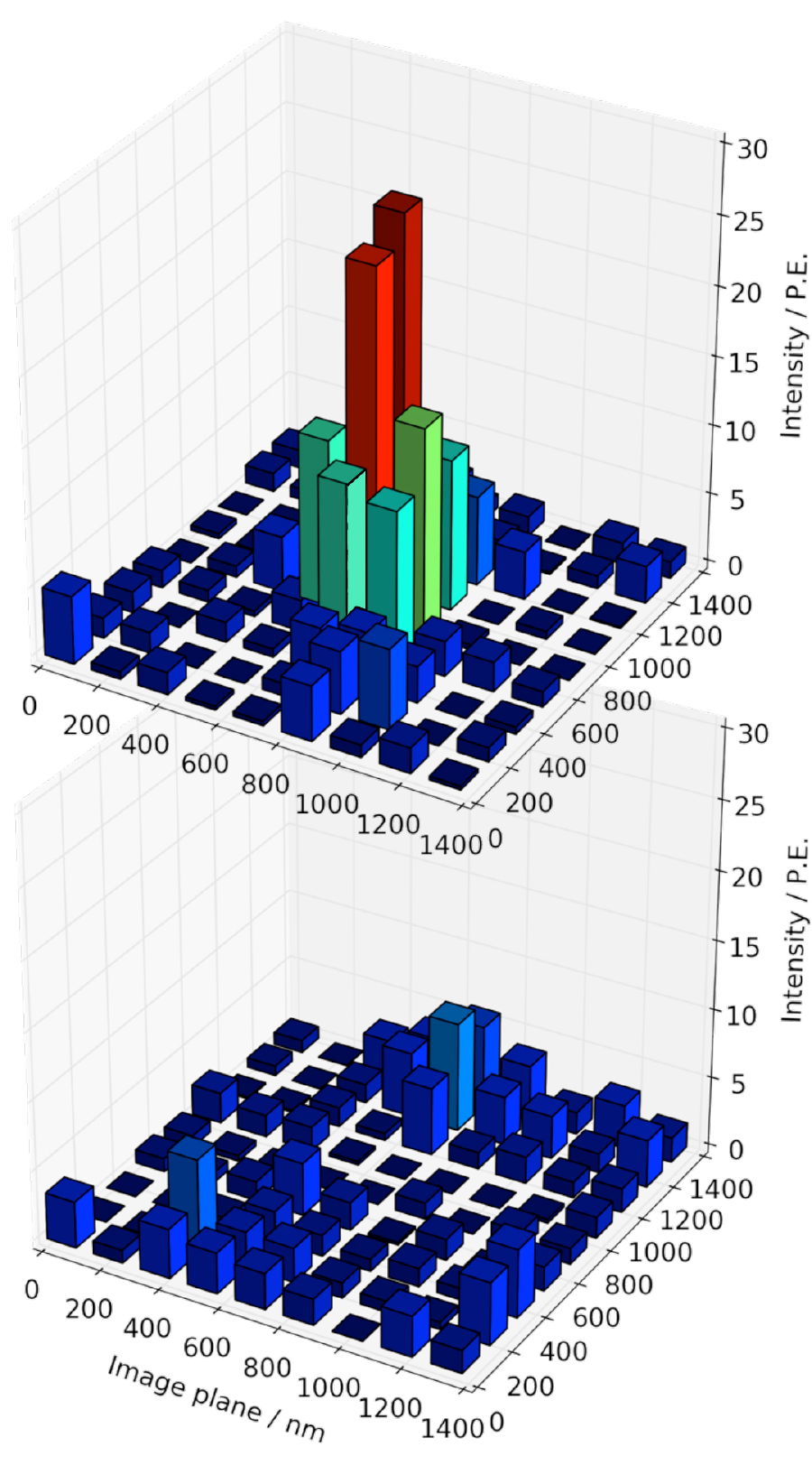} 
\caption{A single Ba$^{++}$ candidate. A fixed region of the CCD camera is shown with 0.5~s exposure before (top) and after (bottom) photo-bleaching transition.\label{fig:legos}}
\end{figure}

In this Letter we describe the resolution of individual Ba$^{++}$ ions on a scanning surface using an SMFI-based sensor (Fig. \ref{fig:legos}), a major step towards barium tagging in HPGXe TPCs.

\section{Apparatus \label{sec:app}}

The SMFI sensor concept developed here uses a thin quartz plate with surface-bound fluorescent indicators, continuously illuminated with excitation light and monitored by an EM-CCD camera.  It is anticipated that such a sensor would form the basis for a Ba$^{++}$ detection system in HPGXe, with ions delivered to a few $\sim$1~mm$^2$ sensing surfaces, first via drift to the cathode and then transversely by RF-carpet \cite{ARAI201456}, a method already demonstrated at large scales \cite{Gehring2016221} and for barium transport in HPGXe \cite{Brunner:2014sfa}.

To demonstrate single Ba$^{++}$ sensitivity we have imaged individual near-surface Ba$^{++}$ ions from dilute barium salt solutions.  We use the technique of through-objective total internal reflection fluorescence (TIRF) microscopy \cite{Burghardt2012}.  In through-objective TIRF, a high numerical aperture objective is used to convert a focused light beam on the back-focal plane (BFP) into parallel rays.  By translating the focal point across the BFP, the emerging parallel rays are brought to a shallow angle, until eventually they totally internally reflect off the lower sample surface.  The evanescent excitation wave, extending only about one wavelength into the sample, causes fluorescence of the near-surface molecules while minimizing deeper fluorescence and scattering from within the sample.  By focusing the objective on the same scanning surface and imaging the fluorescence light, individual fluorescent molecules can be spatially resolved above a dimmer, diffuse background.

\begin{figure}[t]
\begin{center}
\includegraphics[width=0.85\columnwidth]{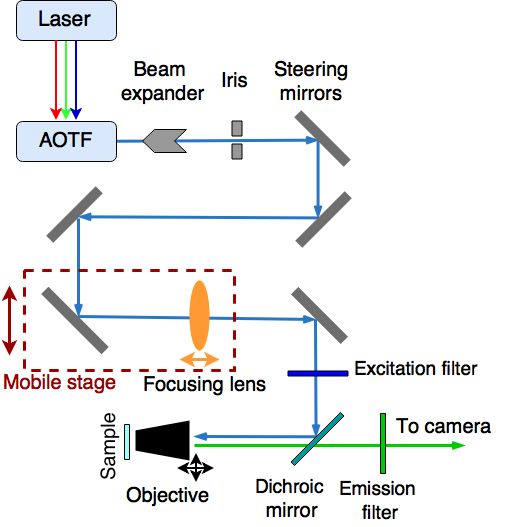} 
\caption{A schematic view of the TIRF system. \label{fig:optics}}
\end{center}
\end{figure}

The apparatus used in this study comprises an Olympus IX-70 inverted fluorescence microscope frame, outfitted with a custom external optical system to implement TIRF imaging. Fig. \ref{fig:optics} shows a schematic view of this system.  The external optics were assembled following the general methods of Ref. \cite{gell2006handbook}. 
The objective is an Olympus 100x 1.4~NA, and was used with Olympus immersion oil ($n=1.518$). Low-fluorescence coverslips  of thickness 0.13~mm serve the combined role of sample substrates and imaging plane. A Hamamatsu ImagEM X2 EM-CCD camera, chosen for its high quantum efficiency, low noise, and high EM gain, was used to record the fluorescence emission.
The excitation source is an NKT Photonics SuperK EXTREME laser, which emits in a wide spectral range of 350-2350~nm.  The laser light is filtered to a tunable band of width $\pm$~1~nm in the visible range via an acousto-optical tunable filter (AOTF). The beam is then expanded, cleaned, and steered via an array of adjustable mirrors through a lens that focuses it onto the objective BFP. Emission and excitation light are separated by a filter cube containing the dichroic mirror and the emission and excitation filters described in \cite{Jones:2016qiq}.  Micrometer stages in the external optical system allow for focusing on the BFP and adjustment of the TIRF ray angle, and the image is brought into focus using sample stage adjustment on the microscope frame.

\begin{figure}[t]
\includegraphics[width=0.95\columnwidth]{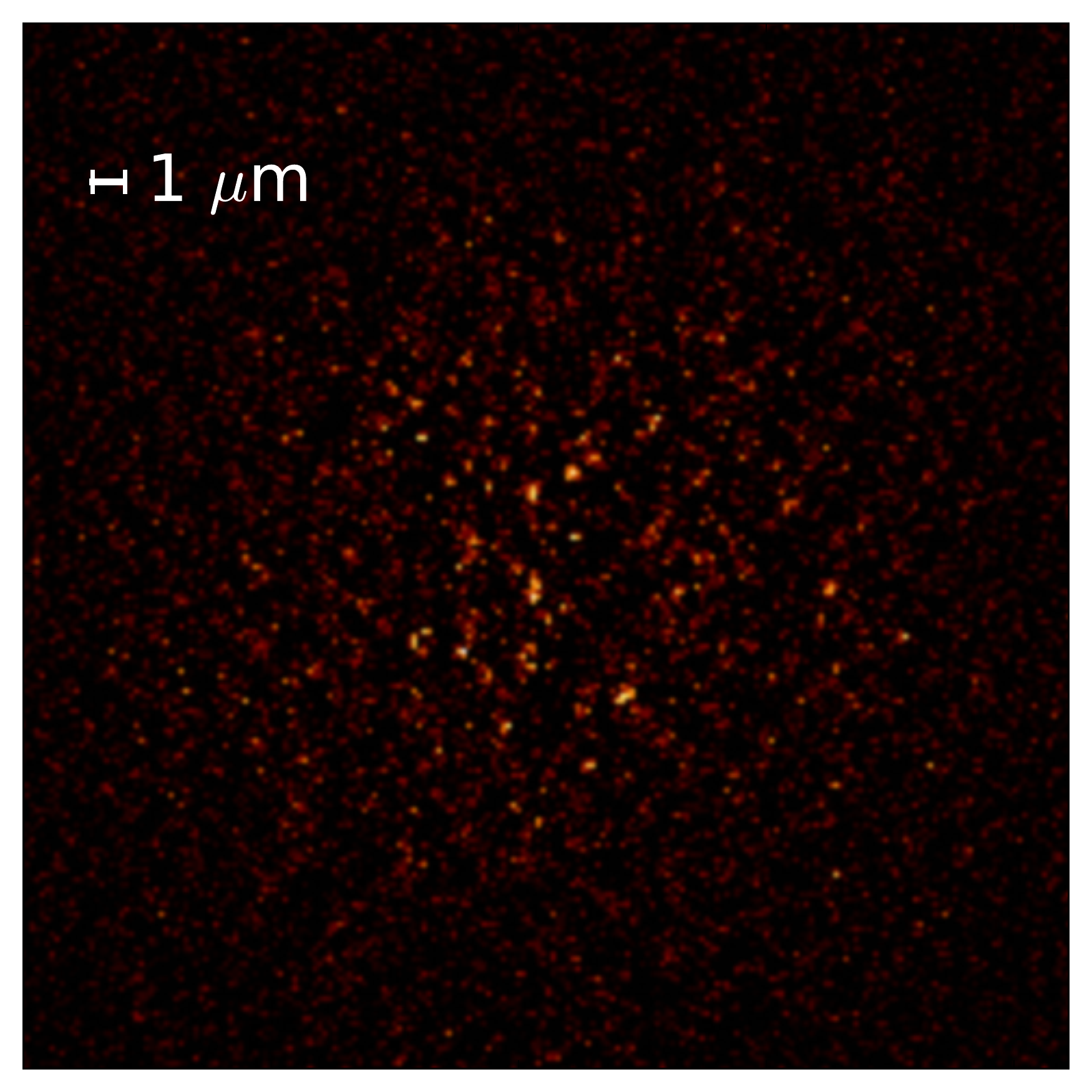} 
\caption{A sample image from the EM-CCD in one of the barium-spiked samples showing both near-surface (bright) and deeper (dim) fluorescent molecules. \label{fig:BaImage}}

\end{figure}

\section{Sample Preparation and Imaging}
The studies presented in this Letter use fluorophores that are immobilized at the sensor surface.  This emulates the conditions in a HPGXe TPC detector, where the ions will drift to the sensor plate and adhere to fluorophores immobilized there. The purpose of this immobilization is to suppress Brownian motion of non-tethered fluorophores in solution to enable spatial resolution. Various methods of immobilization are described in the SMFI literature.  We used a matrix of polyvinyl alcohol (PVA), prepared according to the prescription in Ref. \cite{Stemmer2008}, trapping the fluors but allowing permeability of ions throughout the sample.

Coverslips were inspected for defects and  cleaned by ultrasonic bath in acetone for 30 minutes and then ethanol for 30 minutes. Once clean, they were placed in a vacuum oven at 340~K to bake overnight. A buffer solution (pH~7.2) was established using ACS Ultra Trace water with imidazole and hydrochloric acid. PVA was added (5\% by weight) to form the working buffer and placed in a hot water bath. Once the PVA is fully dissolved it is removed from the bath and set on a stir plate, with water added to reach the target volume. After 30 minutes the solution reaches room temperature and BAPTA \cite{Jones:2016qiq} is added to a concentration of 250~$\mu$M to suppress residual calcium. The SMFI fluor Fluo-3 is then added, to a concentration of 1~nM.  Three background samples were made by placing 50 $\mu$L of solution onto a coverslip then spin coating it at 1800 rpm for 10 seconds. These will be referred to as the ``barium-free'' samples.  Ba$^{++}$ was then added in the form of barium perchlorate salt solution to a target concentration of 500~$\mu$M, to make three ``barium-spiked'' samples. All samples were placed in an oven at 340~K for 4 hours, and then left to cure overnight at room temperature, in order to harden the PVA matrix.

Prior to scanning the samples, the AOTF was set to 488~nm, which is the peak excitation wavelength for Fluo-3.  The measured power  entering the objective was 1~$\mu W$. The external optics were adjusted to place the system into total internal reflection mode using prismatic alignment \cite{fish2009total}, and surface sensitivity was checked with fluorescent micro-spheres.  Each sample was then placed onto the microscope stage and scanned. 

The imaging protocol was to find a 35$\times$35~$\mu$m$^2$ Field of View (FOV) where at least one fluorescent spot was present, and focus the microscope on that spot by minimizing its point spread.  Once focus was achieved, images were taken every 500~ms for 375~s.   Then, a new FOV was found by randomly translating the sample stage and refocusing. This was done for 22 FOVs across the three samples for both barium-free and barium-spiked sets. Notably, because the microscope could not be focused on empty regions, our measurements of the barium-free sample activities are biased towards higher yields, thus imposing a penalty in the numerical significance of our result. Nevertheless, as will be shown, the count of ions in barium-spiked samples is significantly higher than background, demonstrating unambiguous single Ba$^{++}$ ion detection, even with this penalty.

\section{Results}

Figure \ref{fig:BaImage} shows the raw data for one FOV in a barium-spiked sample.  Activity is present from both near-surface (bright) and deeper (dim) fluorophores. An analysis technique was developed to obtain the fluorescence history of the near-surface spots from the raw CCD images.  Only these bright spots would be expected in a surface-based tagging sensor, with the deeper fluorophores here being an artifact of our slide preparation procedure.

\begin{figure}[t]
\includegraphics[width=0.99\columnwidth,height=0.94\columnwidth]{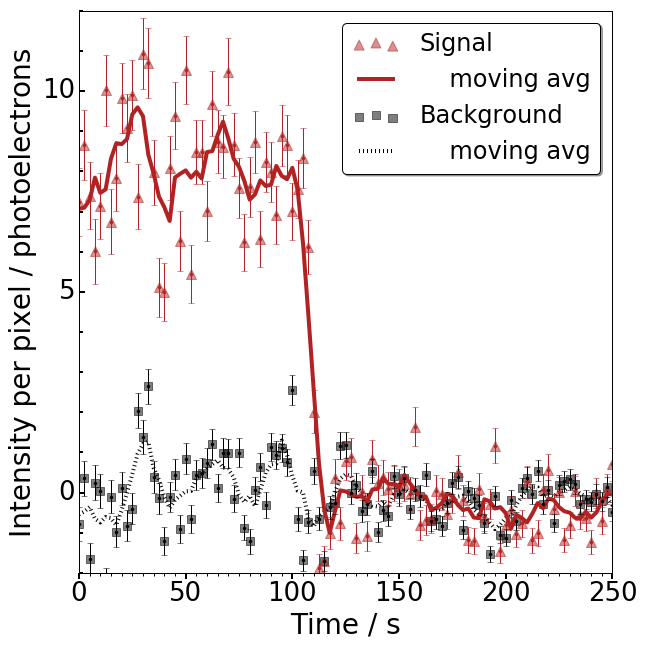} 
\caption{Fluorescence trajectory for one candidate in a barium-spiked sample. ``Signal'' shows the average activity in 5x5 pixels centered on the local maximum. ``Background'' shows the average in the 56 surrounding. The single step photo-bleach is characteristic of single molecule fluorescence. \label{fig:photobleach}}
\end{figure}

The hallmark of single molecule fluorescence is a sudden discrete photo-bleaching transition \cite{Habuchi05072005}.  This occurs when the fluorophore transitions from a fluorescent to a non-fluorescent state, usually via interaction with reactive oxygen species  \cite{thomas2000comparison}.  This discrete transition signifies the presence of a single fluor, rather than a site with multiple fluors contributing. The 375~s scan time is significantly longer than the typical photo-bleaching time of Fluo-3 at this laser power \cite{thomas2000comparison}, so the majority of spots are observed to bleach in our samples. A typical near-surface fluorescence trajectory is shown in Fig.~\ref{fig:photobleach}.  One 0.5~s exposure of this spot directly before the step and one 0.5~s exposure directly after the step are shown in Fig.~\ref{fig:legos}.

Near-surface, photo-bleaching Ba$^{++}$ candidate spots were identified as follows. The images from one FOV were summed and the diffuse background was subtracted. Local maxima 3~$\sigma$ above background fluctuations were identified. The images were analyzed frame by frame and each candidate ion was assigned a 9$\times$9 pixel array centered on that point. The inner 5$\times$5 was summed and taken as the signal, and the outer pixels were summed as a local background reference. This background was fitted to a polynomial function and subtracted, to yield the ``fluorescence trajectory.'' To find  photo-bleaching molecules, each candidate trajectory was fitted against two hypotheses: a straight line and a step function, with $\chi^2$ determined for each. Single photo-bleaching molecules were identified by applying cuts on the step size and the $\chi^2$ difference.

\begin{figure}[t]
\includegraphics[width=0.95\columnwidth]{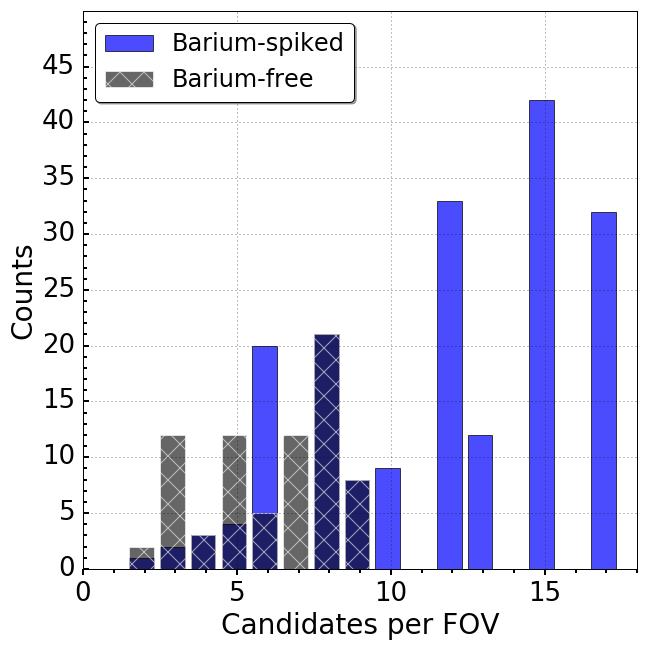} 
\caption{Histogram showing number of ion candidates per FOV in barium-free and barium-spiked samples. Entries are weighted by candidates detected.\label{fig:FOV}}
\end{figure}

The barium-free and barium-spiked samples were analyzed over a total of 22 FOVs each.  Across the barium-free samples, a total of 75 candidates were resolved passing all cuts.  Across the barium-spiked samples, a total of 187 were resolved, representing a statistical excess of $\sim$12.9~$\sigma$.  The actual signal-to-background ratio is larger than is suggested by this comparison, due to the FOV selection bias requiring at least one bright spot to be found in order to focus.  The number of candidates per FOV is shown in Fig.~\ref{fig:FOV}.  In this histogram, each entry is weighted by the number of candidates observed, such that the integral is equal to the total number of candidate ions identified.

Once a single molecule has been identified, its time-integrated intensity profile can be acquired and fitted to a Gaussian shape. We find that the width of this Gaussian is within systematic uncertainty of Abbe’s diffraction limit. Accounting for optical magnification, the localized rms position of the molecule in x and y is $\sim$~2~nm, consistent with the expected super-resolution localization \cite{thompson2002precise}. 

The candidates in the barium-free samples are understood to derive from residual ion contamination in the Ultra Trace water.  Similar background activity was also observed in the bulk studies of  Ref. \cite{Jones:2016qiq}.  In a real barium tagging sensor this background is un-problematic, since the tagging signature is the arrival of a new ion in coincidence with a tagged event. The presence of persistent background fluorophores does not obscure this, unless there are so many that the image is saturated.  The size of the FOV used here and the level of surface contamination suggests that at least $10^5$ Ba$^{++}$ ions will be detectable per sensor before saturation.

\section{Discussion and Outlook}
In this Letter we have demonstrated that SMFI can be used to resolve individual Ba$^{++}$ ions at surfaces via TIRF microscopy.  Ba$^{++}$ ions have been detected above a background of free residual ions at 12.9 $\sigma$ statistical significance, with individual ions spatially resolved and observed to exhibit single-step photo-bleaching trajectories characteristic of single molecules.

An SMFI sensor in a HPGXe TPC will differ from the apparatus described here in a few key ways.  First, the fluorophores will be surface-tethered, and not embedded in a thick sample. Thus, only near-surface bright spots are expected, and offline separation from the deeper background fluors will not be necessary. Second, the target signature will be appearance of a new candidate over a pre-characterized background, coincident in a spatio-temporal region with an $0\nu\beta\beta$ candidate in the TPC. In this case, only the ability to resolve appearance of a new ion is important, and the spatial localization of individual ion candidates demonstrated here shows that many can be recorded on the same sensor before saturation.  Third, the micro-environment around the fluor will be different, being immobilized on a dry surface rather than within a PVA matrix, and this may modify chelation and fluorescence properties of the fluor.  Finally, the extent to which photo-bleaching will be active in a clean HPGXe environment is unknown.  

These issues are now being studied, as the apparatus described here is extended and optimized for SMFI scanning within a xenon gas environment. The chelation and fluorescence of surface-tethered molecules will be characterized by exposing a prepared sensing surface to barium dications from time-of-flight separated beams in xenon gas.  Both commercial and custom fluorophores will be surveyed to obtain optimal performance.  Successful single molecule imaging in this environment would enable identification of Ba$^{++}$ {\em in situ} and in real time, enabling a novel background-free, ton-scale $0\nu\beta\beta$ technology.

\acknowledgments
We thank Yuan Mei and Fernanda Psihas for their comments on this manuscript.  The NEXT Collaboration acknowledges support from the following agencies and institutions: the European Research Council (ERC) under the Advanced Grant 339787-NEXT; the Ministerio de Econom\'ia y Competitividad of Spain under grants FIS2014-53371-C04 and the Severo Ochoa Program SEV-2014-0398; the GVA of Spain under grant PROMETEO/2016/120; the Portuguese FCT and FEDER through the program COMPETE, project PTDC/FIS/103860/2008; the U.S. Department of Energy under contracts number DE-AC02-07CH11359 (Fermi National Accelerator Laboratory) and DE-FG02-13ER42020 (Texas A\&M) and DE-SC0017721 (University of Texas at Arlington); and the University of Texas at Arlington.

\bibliographystyle{apsrev4-1}
\bibliography{smfi}

\begin{thebibliography}{38}%
\makeatletter
\providecommand \@ifxundefined [1]{%
 \@ifx{#1\undefined}
}%
\providecommand \@ifnum [1]{%
 \ifnum #1\expandafter \@firstoftwo
 \else \expandafter \@secondoftwo
 \fi
}%
\providecommand \@ifx [1]{%
 \ifx #1\expandafter \@firstoftwo
 \else \expandafter \@secondoftwo
 \fi
}%
\providecommand \natexlab [1]{#1}%
\providecommand \enquote  [1]{``#1''}%
\providecommand \bibnamefont  [1]{#1}%
\providecommand \bibfnamefont [1]{#1}%
\providecommand \citenamefont [1]{#1}%
\providecommand \href@noop [0]{\@secondoftwo}%
\providecommand \href [0]{\begingroup \@sanitize@url \@href}%
\providecommand \@href[1]{\@@startlink{#1}\@@href}%
\providecommand \@@href[1]{\endgroup#1\@@endlink}%
\providecommand \@sanitize@url [0]{\catcode `\\12\catcode `\$12\catcode
  `\&12\catcode `\#12\catcode `\^12\catcode `\_12\catcode `\%12\relax}%
\providecommand \@@startlink[1]{}%
\providecommand \@@endlink[0]{}%
\providecommand \url  [0]{\begingroup\@sanitize@url \@url }%
\providecommand \@url [1]{\endgroup\@href {#1}{\urlprefix }}%
\providecommand \urlprefix  [0]{URL }%
\providecommand \Eprint [0]{\href }%
\providecommand \doibase [0]{http://dx.doi.org/}%
\providecommand \selectlanguage [0]{\@gobble}%
\providecommand \bibinfo  [0]{\@secondoftwo}%
\providecommand \bibfield  [0]{\@secondoftwo}%
\providecommand \translation [1]{[#1]}%
\providecommand \BibitemOpen [0]{}%
\providecommand \bibitemStop [0]{}%
\providecommand \bibitemNoStop [0]{.\EOS\space}%
\providecommand \EOS [0]{\spacefactor3000\relax}%
\providecommand \BibitemShut  [1]{\csname bibitem#1\endcsname}%
\let\auto@bib@innerbib\@empty
\bibitem [{\citenamefont {Chang}\ and\ \citenamefont
  {Mohapatra}(1985)}]{Chang:1985en}%
  \BibitemOpen
  \bibfield  {author} {\bibinfo {author} {\bibfnamefont {D.}~\bibnamefont
  {Chang}}\ and\ \bibinfo {author} {\bibfnamefont {R.~N.}\ \bibnamefont
  {Mohapatra}},\ }\href {\doibase 10.1103/PhysRevD.32.1248} {\bibfield
  {journal} {\bibinfo  {journal} {Physical Review}\ }\textbf {\bibinfo {volume}
  {D32}},\ \bibinfo {pages} {1248} (\bibinfo {year} {1985})}\BibitemShut
  {NoStop}%
\bibitem [{\citenamefont {Minkowski}(1977)}]{minkowski1977mu}%
  \BibitemOpen
  \bibfield  {author} {\bibinfo {author} {\bibfnamefont {P.}~\bibnamefont
  {Minkowski}},\ }\href@noop {} {\bibfield  {journal} {\bibinfo  {journal}
  {Physics Letters B}\ }\textbf {\bibinfo {volume} {67}},\ \bibinfo {pages}
  {421} (\bibinfo {year} {1977})}\BibitemShut {NoStop}%
\bibitem [{\citenamefont {Gell-Mann}\ \emph {et~al.}(1979)\citenamefont
  {Gell-Mann}, \citenamefont {Ramond},\ and\ \citenamefont
  {Slansky}}]{gell1979ramond}%
  \BibitemOpen
  \bibfield  {author} {\bibinfo {author} {\bibfnamefont {M.}~\bibnamefont
  {Gell-Mann}}, \bibinfo {author} {\bibfnamefont {P.}~\bibnamefont {Ramond}}, \
  and\ \bibinfo {author} {\bibfnamefont {R.}~\bibnamefont {Slansky}},\
  }\href@noop {} {\bibfield  {journal} {\bibinfo  {journal} {Supergravity, ed.
  P. van Nieuwenhuizen and DZ Freedman, North--Holland, Amsterdam}\ } (\bibinfo
  {year} {1979})}\BibitemShut {NoStop}%
\bibitem [{\citenamefont {Yanagida}\ \emph {et~al.}(1979)\citenamefont
  {Yanagida} \emph {et~al.}}]{yanagida1979proceedings}%
  \BibitemOpen
  \bibfield  {author} {\bibinfo {author} {\bibfnamefont {T.}~\bibnamefont
  {Yanagida}} \emph {et~al.},\ }\href@noop {} {\enquote {\bibinfo {title}
  {Proceedings of the workshop on the unified theory and the baryon number in
  the universe},}\ } (\bibinfo {year} {1979})\BibitemShut {NoStop}%
\bibitem [{\citenamefont {Mohapatra}\ and\ \citenamefont
  {Senjanovi{\'c}}(1981)}]{mohapatra1981neutrino}%
  \BibitemOpen
  \bibfield  {author} {\bibinfo {author} {\bibfnamefont {R.~N.}\ \bibnamefont
  {Mohapatra}}\ and\ \bibinfo {author} {\bibfnamefont {G.}~\bibnamefont
  {Senjanovi{\'c}}},\ }\href@noop {} {\bibfield  {journal} {\bibinfo  {journal}
  {Physical Review D}\ }\textbf {\bibinfo {volume} {23}},\ \bibinfo {pages}
  {165} (\bibinfo {year} {1981})}\BibitemShut {NoStop}%
\bibitem [{\citenamefont {Fukugita}\ and\ \citenamefont
  {Yanagida}(1986)}]{Fukugita:1986hr}%
  \BibitemOpen
  \bibfield  {author} {\bibinfo {author} {\bibfnamefont {M.}~\bibnamefont
  {Fukugita}}\ and\ \bibinfo {author} {\bibfnamefont {T.}~\bibnamefont
  {Yanagida}},\ }\href {\doibase 10.1016/0370-2693(86)91126-3} {\bibfield
  {journal} {\bibinfo  {journal} {Phys. Lett.}\ }\textbf {\bibinfo {volume}
  {B174}},\ \bibinfo {pages} {45} (\bibinfo {year} {1986})}\BibitemShut
  {NoStop}%
\bibitem [{\citenamefont {Ostrovskiy}\ and\ \citenamefont
  {O'Sullivan}(2016)}]{Ostrovskiy:2016uyx}%
  \BibitemOpen
  \bibfield  {author} {\bibinfo {author} {\bibfnamefont {I.}~\bibnamefont
  {Ostrovskiy}}\ and\ \bibinfo {author} {\bibfnamefont {K.}~\bibnamefont
  {O'Sullivan}},\ }\href {\doibase 10.1142/S0217732316920048,
  10.1142/S0217732316300172} {\bibfield  {journal} {\bibinfo  {journal} {Mod.
  Phys. Lett.}\ }\textbf {\bibinfo {volume} {A31}},\ \bibinfo {pages} {1630017}
  (\bibinfo {year} {2016})},\ \bibinfo {note} {[Erratum: Mod. Phys.
  Lett.A31,no.23,1692004(2016)]},\ \Eprint {http://arxiv.org/abs/1605.00631}
  {arXiv:1605.00631 [hep-ex]} \BibitemShut {NoStop}%
\bibitem [{\citenamefont {Dell'Oro}\ \emph {et~al.}(2016)\citenamefont
  {Dell'Oro}, \citenamefont {Marcocci}, \citenamefont {Viel},\ and\
  \citenamefont {Vissani}}]{DellOro:2016tmg}%
  \BibitemOpen
  \bibfield  {author} {\bibinfo {author} {\bibfnamefont {S.}~\bibnamefont
  {Dell'Oro}}, \bibinfo {author} {\bibfnamefont {S.}~\bibnamefont {Marcocci}},
  \bibinfo {author} {\bibfnamefont {M.}~\bibnamefont {Viel}}, \ and\ \bibinfo
  {author} {\bibfnamefont {F.}~\bibnamefont {Vissani}},\ }\href {\doibase
  10.1155/2016/2162659} {\bibfield  {journal} {\bibinfo  {journal} {Adv. High
  Energy Phys.}\ }\textbf {\bibinfo {volume} {2016}},\ \bibinfo {pages}
  {2162659} (\bibinfo {year} {2016})},\ \Eprint
  {http://arxiv.org/abs/1601.07512} {arXiv:1601.07512 [hep-ph]} \BibitemShut
  {NoStop}%
\bibitem [{\citenamefont {Gomez-Cadenas}\ \emph
  {et~al.}(2011{\natexlab{a}})\citenamefont {Gomez-Cadenas} \emph
  {et~al.}}]{GomezCadenas:2010gs}%
  \BibitemOpen
  \bibfield  {author} {\bibinfo {author} {\bibfnamefont {J.~J.}\ \bibnamefont
  {Gomez-Cadenas}} \emph {et~al.},\ }\href {\doibase
  10.1088/1475-7516/2011/06/007} {\bibfield  {journal} {\bibinfo  {journal}
  {JCAP}\ }\textbf {\bibinfo {volume} {1106}},\ \bibinfo {pages} {007}
  (\bibinfo {year} {2011}{\natexlab{a}})},\ \Eprint
  {http://arxiv.org/abs/1010.5112} {arXiv:1010.5112 [hep-ex]} \BibitemShut
  {NoStop}%
\bibitem [{\citenamefont {Gomez-Cadenas}\ \emph
  {et~al.}(2011{\natexlab{b}})\citenamefont {Gomez-Cadenas}, \citenamefont
  {Martin-Albo}, \citenamefont {Mezzetto}, \citenamefont {Monrabal},\ and\
  \citenamefont {Sorel}}]{gomez2011search}%
  \BibitemOpen
  \bibfield  {author} {\bibinfo {author} {\bibfnamefont {J.}~\bibnamefont
  {Gomez-Cadenas}}, \bibinfo {author} {\bibfnamefont {J.}~\bibnamefont
  {Martin-Albo}}, \bibinfo {author} {\bibfnamefont {M.}~\bibnamefont
  {Mezzetto}}, \bibinfo {author} {\bibfnamefont {F.}~\bibnamefont {Monrabal}},
  \ and\ \bibinfo {author} {\bibfnamefont {M.}~\bibnamefont {Sorel}},\
  }\href@noop {} {\bibfield  {journal} {\bibinfo  {journal} {arXiv preprint
  arXiv:1109.5515}\ } (\bibinfo {year} {2011}{\natexlab{b}})}\BibitemShut
  {NoStop}%
\bibitem [{\citenamefont {Aseev}\ \emph {et~al.}(2011)\citenamefont {Aseev}
  \emph {et~al.}}]{Aseev:2011dq}%
  \BibitemOpen
  \bibfield  {author} {\bibinfo {author} {\bibfnamefont {V.~N.}\ \bibnamefont
  {Aseev}} \emph {et~al.} (\bibinfo {collaboration} {Troitsk}),\ }\href
  {\doibase 10.1103/PhysRevD.84.112003} {\bibfield  {journal} {\bibinfo
  {journal} {Phys. Rev.}\ }\textbf {\bibinfo {volume} {D84}},\ \bibinfo {pages}
  {112003} (\bibinfo {year} {2011})},\ \Eprint {http://arxiv.org/abs/1108.5034}
  {arXiv:1108.5034 [hep-ex]} \BibitemShut {NoStop}%
\bibitem [{\citenamefont {Ade}\ \emph {et~al.}(2016)\citenamefont {Ade} \emph
  {et~al.}}]{Ade:2015xua}%
  \BibitemOpen
  \bibfield  {author} {\bibinfo {author} {\bibfnamefont {P.~A.~R.}\
  \bibnamefont {Ade}} \emph {et~al.} (\bibinfo {collaboration} {Planck}),\
  }\href {\doibase 10.1051/0004-6361/201525830} {\bibfield  {journal} {\bibinfo
   {journal} {Astron. Astrophys.}\ }\textbf {\bibinfo {volume} {594}},\
  \bibinfo {pages} {A13} (\bibinfo {year} {2016})},\ \Eprint
  {http://arxiv.org/abs/1502.01589} {arXiv:1502.01589 [astro-ph.CO]}
  \BibitemShut {NoStop}%
\bibitem [{\citenamefont {Gonzalez-Garcia}\ \emph {et~al.}(2016)\citenamefont
  {Gonzalez-Garcia}, \citenamefont {Maltoni},\ and\ \citenamefont
  {Schwetz}}]{Gonzalez-Garcia:2015qrr}%
  \BibitemOpen
  \bibfield  {author} {\bibinfo {author} {\bibfnamefont {M.~C.}\ \bibnamefont
  {Gonzalez-Garcia}}, \bibinfo {author} {\bibfnamefont {M.}~\bibnamefont
  {Maltoni}}, \ and\ \bibinfo {author} {\bibfnamefont {T.}~\bibnamefont
  {Schwetz}},\ }\href {\doibase 10.1016/j.nuclphysb.2016.02.033} {\bibfield
  {journal} {\bibinfo  {journal} {Nucl. Phys.}\ }\textbf {\bibinfo {volume}
  {B908}},\ \bibinfo {pages} {199} (\bibinfo {year} {2016})},\ \Eprint
  {http://arxiv.org/abs/1512.06856} {arXiv:1512.06856 [hep-ph]} \BibitemShut
  {NoStop}%
\bibitem [{\citenamefont {Alvarez}\ \emph {et~al.}(2012)\citenamefont {Alvarez}
  \emph {et~al.}}]{Alvarez:2012flf}%
  \BibitemOpen
  \bibfield  {author} {\bibinfo {author} {\bibfnamefont {V.}~\bibnamefont
  {Alvarez}} \emph {et~al.} (\bibinfo {collaboration} {NEXT}),\ }\href
  {\doibase 10.1088/1748-0221/7/06/T06001} {\bibfield  {journal} {\bibinfo
  {journal} {JINST}\ }\textbf {\bibinfo {volume} {7}},\ \bibinfo {pages}
  {T06001} (\bibinfo {year} {2012})},\ \Eprint {http://arxiv.org/abs/1202.0721}
  {arXiv:1202.0721 [physics.ins-det]} \BibitemShut {NoStop}%
\bibitem [{\citenamefont {Auger}\ \emph {et~al.}(2012)\citenamefont {Auger}
  \emph {et~al.}}]{Auger:2012ar}%
  \BibitemOpen
  \bibfield  {author} {\bibinfo {author} {\bibfnamefont {M.}~\bibnamefont
  {Auger}} \emph {et~al.} (\bibinfo {collaboration} {EXO-200}),\ }\href
  {\doibase 10.1103/PhysRevLett.109.032505} {\bibfield  {journal} {\bibinfo
  {journal} {Phys. Rev. Lett.}\ }\textbf {\bibinfo {volume} {109}},\ \bibinfo
  {pages} {032505} (\bibinfo {year} {2012})},\ \Eprint
  {http://arxiv.org/abs/1205.5608} {arXiv:1205.5608 [hep-ex]} \BibitemShut
  {NoStop}%
\bibitem [{\citenamefont {Moe}(1991)}]{Moe:1991ik}%
  \BibitemOpen
  \bibfield  {author} {\bibinfo {author} {\bibfnamefont {M.~K.}\ \bibnamefont
  {Moe}},\ }\href {\doibase 10.1103/PhysRevC.44.931} {\bibfield  {journal}
  {\bibinfo  {journal} {Physical Review}\ }\textbf {\bibinfo {volume} {C44}},\
  \bibinfo {pages} {931} (\bibinfo {year} {1991})}\BibitemShut {NoStop}%
\bibitem [{\citenamefont {Danilov}\ \emph {et~al.}(2000)\citenamefont {Danilov}
  \emph {et~al.}}]{Danilov:2000pp}%
  \BibitemOpen
  \bibfield  {author} {\bibinfo {author} {\bibfnamefont {M.}~\bibnamefont
  {Danilov}} \emph {et~al.},\ }\bibfield  {booktitle} {\emph {\bibinfo
  {booktitle} {{Low energy solar neutrino detection. Proceedings, 2nd
  International Workshop, LoWNu2, Tokyo, Japan, December 4-5, 2000}}},\ }\href
  {\doibase 10.1016/S0370-2693(00)00404-4} {\bibfield  {journal} {\bibinfo
  {journal} {Physics Letters}\ }\textbf {\bibinfo {volume} {B480}},\ \bibinfo
  {pages} {12} (\bibinfo {year} {2000})},\ \Eprint
  {http://arxiv.org/abs/hep-ex/0002003} {arXiv:hep-ex/0002003 [hep-ex]}
  \BibitemShut {NoStop}%
\bibitem [{\citenamefont {Mong}\ \emph {et~al.}(2015)\citenamefont {Mong} \emph
  {et~al.}}]{Mong:2014iya}%
  \BibitemOpen
  \bibfield  {author} {\bibinfo {author} {\bibfnamefont {B.}~\bibnamefont
  {Mong}} \emph {et~al.},\ }\href {\doibase 10.1103/PhysRevA.91.022505}
  {\bibfield  {journal} {\bibinfo  {journal} {Physical Review}\ }\textbf
  {\bibinfo {volume} {A91}},\ \bibinfo {pages} {022505} (\bibinfo {year}
  {2015})},\ \Eprint {http://arxiv.org/abs/1410.2624} {arXiv:1410.2624
  [physics.atom-ph]} \BibitemShut {NoStop}%
\bibitem [{\citenamefont {Rollin}(2011)}]{Rollin:2011gla}%
  \BibitemOpen
  \bibfield  {author} {\bibinfo {author} {\bibfnamefont {E.}~\bibnamefont
  {Rollin}},\ }\emph {\bibinfo {title} {{Barium Ion Extraction and
  Identification from Laser Induced Fluorescence in Gas for the Enriched Xenon
  Observatory}}},\ \href
  {https://curve.carleton.ca/system/files/theses/31212.pdf} {Ph.D. thesis},\
  \bibinfo  {school} {Carleton U.} (\bibinfo {year} {2011})\BibitemShut
  {NoStop}%
\bibitem [{\citenamefont {Brunner}\ \emph {et~al.}(2015)\citenamefont {Brunner}
  \emph {et~al.}}]{Brunner:2014sfa}%
  \BibitemOpen
  \bibfield  {author} {\bibinfo {author} {\bibfnamefont {T.}~\bibnamefont
  {Brunner}} \emph {et~al.},\ }\href {\doibase 10.1016/j.ijms.2015.01.003}
  {\bibfield  {journal} {\bibinfo  {journal} {International Journal of Mass
  Spectrometry}\ }\textbf {\bibinfo {volume} {379}},\ \bibinfo {pages} {110}
  (\bibinfo {year} {2015})},\ \Eprint {http://arxiv.org/abs/1412.1144}
  {arXiv:1412.1144 [physics.ins-det]} \BibitemShut {NoStop}%
\bibitem [{\citenamefont {Flatt}\ \emph {et~al.}(2007)\citenamefont {Flatt}
  \emph {et~al.}}]{Flatt:2007aa}%
  \BibitemOpen
  \bibfield  {author} {\bibinfo {author} {\bibfnamefont {B.}~\bibnamefont
  {Flatt}} \emph {et~al.},\ }\href {\doibase 10.1016/j.nima.2007.05.123}
  {\bibfield  {journal} {\bibinfo  {journal} {Nuclear Instruments and Methods
  in Physics Research Section A: Accelerators, Spectrometers, Detectors and
  Associated Equipment}\ }\textbf {\bibinfo {volume} {A578}},\ \bibinfo {pages}
  {399} (\bibinfo {year} {2007})},\ \Eprint {http://arxiv.org/abs/0704.1646}
  {arXiv:0704.1646 [physics.ins-det]} \BibitemShut {NoStop}%
\bibitem [{\citenamefont {Sinclair}\ \emph {et~al.}(2011)\citenamefont
  {Sinclair} \emph {et~al.}}]{Sinclair:2011zz}%
  \BibitemOpen
  \bibfield  {author} {\bibinfo {author} {\bibfnamefont {D.}~\bibnamefont
  {Sinclair}} \emph {et~al.},\ }\bibfield  {booktitle} {\emph {\bibinfo
  {booktitle} {{Proceedings, 5th Symposium on Large TPCs for Low Energy Rare
  Event Detection and Workshop on Neutrinos from Supernovae}}},\ }\href
  {\doibase 10.1088/1742-6596/309/1/012005} {\bibfield  {journal} {\bibinfo
  {journal} {Journal of Physics Conference Series}\ }\textbf {\bibinfo {volume}
  {309}},\ \bibinfo {pages} {012005} (\bibinfo {year} {2011})}\BibitemShut
  {NoStop}%
\bibitem [{\citenamefont {Lu}\ and\ \citenamefont {Paige}(2007)}]{Lu2007}%
  \BibitemOpen
  \bibfield  {author} {\bibinfo {author} {\bibfnamefont {Y.}~\bibnamefont
  {Lu}}\ and\ \bibinfo {author} {\bibfnamefont {M.~F.}\ \bibnamefont {Paige}},\
  }\href {\doibase 10.1007/s10895-007-0185-1} {\bibfield  {journal} {\bibinfo
  {journal} {Journal of Fluorescence}\ }\textbf {\bibinfo {volume} {17}},\
  \bibinfo {pages} {739} (\bibinfo {year} {2007})}\BibitemShut {NoStop}%
\bibitem [{\citenamefont {Stuurman}\ and\ \citenamefont
  {Vale}(2006)}]{stuurman2006imaging}%
  \BibitemOpen
  \bibfield  {author} {\bibinfo {author} {\bibfnamefont {N.}~\bibnamefont
  {Stuurman}}\ and\ \bibinfo {author} {\bibfnamefont {R.}~\bibnamefont
  {Vale}},\ }\href@noop {} {\enquote {\bibinfo {title} {Imaging single
  molecules using total internal reflection fluorescence microscopy},}\ }
  (\bibinfo {year} {2006})\BibitemShut {NoStop}%
\bibitem [{\citenamefont {{\'A}lvarez}\ \emph {et~al.}(2013)\citenamefont
  {{\'A}lvarez}, \citenamefont {Borges}, \citenamefont {C{\'a}rcel},
  \citenamefont {Castel}, \citenamefont {Cebri{\'a}n}, \citenamefont {Cervera},
  \citenamefont {Conde}, \citenamefont {Dafni}, \citenamefont {Dias},
  \citenamefont {D{\'\i}az}, \citenamefont {Egorov}, \citenamefont {Esteve},
  \citenamefont {Evtoukhovitch}, \citenamefont {Fernandes}, \citenamefont
  {Ferrario}, \citenamefont {Ferreira}, \citenamefont {Freitas}, \citenamefont
  {Gehman}, \citenamefont {Gil}, \citenamefont {Goldschmidt}, \citenamefont
  {G{\'o}mez}, \citenamefont {G{\'o}mez-Cadenas}, \citenamefont
  {Gonz{\'a}lez-D{\'\i}az}, \citenamefont {Guti{\'e}rrez}, \citenamefont
  {Hauptman}, \citenamefont {Morata}, \citenamefont {Herrera}, \citenamefont
  {Iguaz}, \citenamefont {Irastorza}, \citenamefont {Jinete}, \citenamefont
  {Labarga}, \citenamefont {Laing}, \citenamefont {Liubarsky}, \citenamefont
  {Lopes}, \citenamefont {Lorca}, \citenamefont {Losada}, \citenamefont
  {Luz{\'o}n}, \citenamefont {Mar{\'\i}}, \citenamefont {Mart{\'\i}n-Albo},
  \citenamefont {Mart{\'\i}nez}, \citenamefont {Miller}, \citenamefont
  {Moiseenko}, \citenamefont {Monrabal}, \citenamefont {Monteiro},
  \citenamefont {Mora}, \citenamefont {Moutinho}, \citenamefont {Vidal},
  \citenamefont {da~Luz}, \citenamefont {Navarro}, \citenamefont {Nebot},
  \citenamefont {Nygren}, \citenamefont {Oliveira}, \citenamefont {Palma},
  \citenamefont {P{\'e}rez}, \citenamefont {Aparicio}, \citenamefont {Renner},
  \citenamefont {Ripoll}, \citenamefont {Rodr{\'\i}guez}, \citenamefont
  {Rodr{\'\i}guez}, \citenamefont {Santos}, \citenamefont {dos Santos},
  \citenamefont {Segui}, \citenamefont {Serra}, \citenamefont {Shuman},
  \citenamefont {Sim{\'o}n}, \citenamefont {Sofka}, \citenamefont {Sorel},
  \citenamefont {Toledo}, \citenamefont {Tom{\'a}s}, \citenamefont {Torrent},
  \citenamefont {Tsamalaidze}, \citenamefont {V{\'a}zquez}, \citenamefont
  {Veloso}, \citenamefont {Villar}, \citenamefont {Webb}, \citenamefont
  {White},\ and\ \citenamefont {Yahlali}}]{Alvarez:2012yxw}%
  \BibitemOpen
  \bibfield  {author} {\bibinfo {author} {\bibfnamefont {V.}~\bibnamefont
  {{\'A}lvarez}}, \bibinfo {author} {\bibfnamefont {F.~I. G.~M.}\ \bibnamefont
  {Borges}}, \bibinfo {author} {\bibfnamefont {S.}~\bibnamefont {C{\'a}rcel}},
  \bibinfo {author} {\bibfnamefont {J.}~\bibnamefont {Castel}}, \bibinfo
  {author} {\bibfnamefont {S.}~\bibnamefont {Cebri{\'a}n}}, \bibinfo {author}
  {\bibfnamefont {A.}~\bibnamefont {Cervera}}, \bibinfo {author} {\bibfnamefont
  {C.~A.~N.}\ \bibnamefont {Conde}}, \bibinfo {author} {\bibfnamefont
  {T.}~\bibnamefont {Dafni}}, \bibinfo {author} {\bibfnamefont {T.~H. V.~T.}\
  \bibnamefont {Dias}}, \bibinfo {author} {\bibfnamefont {J.}~\bibnamefont
  {D{\'\i}az}}, \bibinfo {author} {\bibfnamefont {M.}~\bibnamefont {Egorov}},
  \bibinfo {author} {\bibfnamefont {R.}~\bibnamefont {Esteve}}, \bibinfo
  {author} {\bibfnamefont {P.}~\bibnamefont {Evtoukhovitch}}, \bibinfo {author}
  {\bibfnamefont {L.~M.~P.}\ \bibnamefont {Fernandes}}, \bibinfo {author}
  {\bibfnamefont {P.}~\bibnamefont {Ferrario}}, \bibinfo {author}
  {\bibfnamefont {A.~L.}\ \bibnamefont {Ferreira}}, \bibinfo {author}
  {\bibfnamefont {E.~D.~C.}\ \bibnamefont {Freitas}}, \bibinfo {author}
  {\bibfnamefont {V.~M.}\ \bibnamefont {Gehman}}, \bibinfo {author}
  {\bibfnamefont {A.}~\bibnamefont {Gil}}, \bibinfo {author} {\bibfnamefont
  {A.}~\bibnamefont {Goldschmidt}}, \bibinfo {author} {\bibfnamefont
  {H.}~\bibnamefont {G{\'o}mez}}, \bibinfo {author} {\bibfnamefont {J.~J.}\
  \bibnamefont {G{\'o}mez-Cadenas}}, \bibinfo {author} {\bibfnamefont
  {D.}~\bibnamefont {Gonz{\'a}lez-D{\'\i}az}}, \bibinfo {author} {\bibfnamefont
  {R.~M.}\ \bibnamefont {Guti{\'e}rrez}}, \bibinfo {author} {\bibfnamefont
  {J.}~\bibnamefont {Hauptman}}, \bibinfo {author} {\bibfnamefont {J.~A.~H.}\
  \bibnamefont {Morata}}, \bibinfo {author} {\bibfnamefont {D.~C.}\
  \bibnamefont {Herrera}}, \bibinfo {author} {\bibfnamefont {F.~J.}\
  \bibnamefont {Iguaz}}, \bibinfo {author} {\bibfnamefont {I.~G.}\ \bibnamefont
  {Irastorza}}, \bibinfo {author} {\bibfnamefont {M.~A.}\ \bibnamefont
  {Jinete}}, \bibinfo {author} {\bibfnamefont {L.}~\bibnamefont {Labarga}},
  \bibinfo {author} {\bibfnamefont {A.}~\bibnamefont {Laing}}, \bibinfo
  {author} {\bibfnamefont {I.}~\bibnamefont {Liubarsky}}, \bibinfo {author}
  {\bibfnamefont {J.~A.~M.}\ \bibnamefont {Lopes}}, \bibinfo {author}
  {\bibfnamefont {D.}~\bibnamefont {Lorca}}, \bibinfo {author} {\bibfnamefont
  {M.}~\bibnamefont {Losada}}, \bibinfo {author} {\bibfnamefont
  {G.}~\bibnamefont {Luz{\'o}n}}, \bibinfo {author} {\bibfnamefont
  {A.}~\bibnamefont {Mar{\'\i}}}, \bibinfo {author} {\bibfnamefont
  {J.}~\bibnamefont {Mart{\'\i}n-Albo}}, \bibinfo {author} {\bibfnamefont
  {A.}~\bibnamefont {Mart{\'\i}nez}}, \bibinfo {author} {\bibfnamefont
  {T.}~\bibnamefont {Miller}}, \bibinfo {author} {\bibfnamefont
  {A.}~\bibnamefont {Moiseenko}}, \bibinfo {author} {\bibfnamefont
  {F.}~\bibnamefont {Monrabal}}, \bibinfo {author} {\bibfnamefont {C.~M.~B.}\
  \bibnamefont {Monteiro}}, \bibinfo {author} {\bibfnamefont {F.~J.}\
  \bibnamefont {Mora}}, \bibinfo {author} {\bibfnamefont {L.~M.}\ \bibnamefont
  {Moutinho}}, \bibinfo {author} {\bibfnamefont {J.~M.}\ \bibnamefont {Vidal}},
  \bibinfo {author} {\bibfnamefont {H.~N.}\ \bibnamefont {da~Luz}}, \bibinfo
  {author} {\bibfnamefont {G.}~\bibnamefont {Navarro}}, \bibinfo {author}
  {\bibfnamefont {M.}~\bibnamefont {Nebot}}, \bibinfo {author} {\bibfnamefont
  {D.}~\bibnamefont {Nygren}}, \bibinfo {author} {\bibfnamefont {C.~A.~B.}\
  \bibnamefont {Oliveira}}, \bibinfo {author} {\bibfnamefont {R.}~\bibnamefont
  {Palma}}, \bibinfo {author} {\bibfnamefont {J.}~\bibnamefont {P{\'e}rez}},
  \bibinfo {author} {\bibfnamefont {J.~L.~P.}\ \bibnamefont {Aparicio}},
  \bibinfo {author} {\bibfnamefont {J.}~\bibnamefont {Renner}}, \bibinfo
  {author} {\bibfnamefont {L.}~\bibnamefont {Ripoll}}, \bibinfo {author}
  {\bibfnamefont {A.}~\bibnamefont {Rodr{\'\i}guez}}, \bibinfo {author}
  {\bibfnamefont {J.}~\bibnamefont {Rodr{\'\i}guez}}, \bibinfo {author}
  {\bibfnamefont {F.~P.}\ \bibnamefont {Santos}}, \bibinfo {author}
  {\bibfnamefont {J.~M.~F.}\ \bibnamefont {dos Santos}}, \bibinfo {author}
  {\bibfnamefont {L.}~\bibnamefont {Segui}}, \bibinfo {author} {\bibfnamefont
  {L.}~\bibnamefont {Serra}}, \bibinfo {author} {\bibfnamefont
  {D.}~\bibnamefont {Shuman}}, \bibinfo {author} {\bibfnamefont
  {A.}~\bibnamefont {Sim{\'o}n}}, \bibinfo {author} {\bibfnamefont
  {C.}~\bibnamefont {Sofka}}, \bibinfo {author} {\bibfnamefont
  {M.}~\bibnamefont {Sorel}}, \bibinfo {author} {\bibfnamefont {J.~F.}\
  \bibnamefont {Toledo}}, \bibinfo {author} {\bibfnamefont {A.}~\bibnamefont
  {Tom{\'a}s}}, \bibinfo {author} {\bibfnamefont {J.}~\bibnamefont {Torrent}},
  \bibinfo {author} {\bibfnamefont {Z.}~\bibnamefont {Tsamalaidze}}, \bibinfo
  {author} {\bibfnamefont {D.}~\bibnamefont {V{\'a}zquez}}, \bibinfo {author}
  {\bibfnamefont {J.~F. C.~A.}\ \bibnamefont {Veloso}}, \bibinfo {author}
  {\bibfnamefont {J.~A.}\ \bibnamefont {Villar}}, \bibinfo {author}
  {\bibfnamefont {R.}~\bibnamefont {Webb}}, \bibinfo {author} {\bibfnamefont
  {J.~T.}\ \bibnamefont {White}}, \ and\ \bibinfo {author} {\bibfnamefont
  {N.}~\bibnamefont {Yahlali}} (\bibinfo {collaboration} {NEXT}),\ }\href
  {\doibase 10.1016/j.nima.2012.12.123} {\bibfield  {journal} {\bibinfo
  {journal} {Nuclear Instruments and Methods in Physics Research Section A:
  Accelerators, Spectrometers, Detectors and Associated Equipment}\ }\textbf
  {\bibinfo {volume} {A708}},\ \bibinfo {pages} {101} (\bibinfo {year}
  {2013})},\ \Eprint {http://arxiv.org/abs/1211.4474} {arXiv:1211.4474
  [physics.ins-det]} \BibitemShut {NoStop}%
\bibitem [{\citenamefont {Green}(1957)}]{PhysRev.107.1646}%
  \BibitemOpen
  \bibfield  {author} {\bibinfo {author} {\bibfnamefont {A.~E.~S.}\
  \bibnamefont {Green}},\ }\href {\doibase 10.1103/PhysRev.107.1646} {\bibfield
   {journal} {\bibinfo  {journal} {Physical Review}\ }\textbf {\bibinfo
  {volume} {107}},\ \bibinfo {pages} {1646} (\bibinfo {year}
  {1957})}\BibitemShut {NoStop}%
\bibitem [{\citenamefont {Albert}\ \emph {et~al.}(2015)\citenamefont {Albert}
  \emph {et~al.}}]{PhysRevC.92.045504}%
  \BibitemOpen
  \bibfield  {author} {\bibinfo {author} {\bibfnamefont {J.~B.}\ \bibnamefont
  {Albert}} \emph {et~al.} (\bibinfo {collaboration} {EXO-200 Collaboration}),\
  }\href {\doibase 10.1103/PhysRevC.92.045504} {\bibfield  {journal} {\bibinfo
  {journal} {Phys. Rev. C}\ }\textbf {\bibinfo {volume} {92}},\ \bibinfo
  {pages} {045504} (\bibinfo {year} {2015})}\BibitemShut {NoStop}%
\bibitem [{\citenamefont {{Bolotnikov}}\ and\ \citenamefont
  {{Ramsey}}(1997)}]{1997NIMPA.396..360B}%
  \BibitemOpen
  \bibfield  {author} {\bibinfo {author} {\bibfnamefont {A.}~\bibnamefont
  {{Bolotnikov}}}\ and\ \bibinfo {author} {\bibfnamefont {B.}~\bibnamefont
  {{Ramsey}}},\ }\href {\doibase 10.1016/S0168-9002(97)00784-5} {\bibfield
  {journal} {\bibinfo  {journal} {Nuclear Instruments and Methods in Physics
  Research A}\ }\textbf {\bibinfo {volume} {396}},\ \bibinfo {pages} {360}
  (\bibinfo {year} {1997})}\BibitemShut {NoStop}%
\bibitem [{\citenamefont {Jones}\ \emph {et~al.}(2016)\citenamefont {Jones},
  \citenamefont {McDonald},\ and\ \citenamefont {Nygren}}]{Jones:2016qiq}%
  \BibitemOpen
  \bibfield  {author} {\bibinfo {author} {\bibfnamefont {B.~J.~P.}\
  \bibnamefont {Jones}}, \bibinfo {author} {\bibfnamefont {A.~D.}\ \bibnamefont
  {McDonald}}, \ and\ \bibinfo {author} {\bibfnamefont {D.~R.}\ \bibnamefont
  {Nygren}},\ }\href@noop {} {\  (\bibinfo {year} {2016})},\ \Eprint
  {http://arxiv.org/abs/1609.04019} {arXiv:1609.04019 [physics.ins-det]}
  \BibitemShut {NoStop}%
\bibitem [{\citenamefont {Arai}\ \emph {et~al.}(2014)\citenamefont {Arai} \emph
  {et~al.}}]{ARAI201456}%
  \BibitemOpen
  \bibfield  {author} {\bibinfo {author} {\bibfnamefont {F.}~\bibnamefont
  {Arai}} \emph {et~al.},\ }\href {\doibase
  http://dx.doi.org/10.1016/j.ijms.2014.01.005} {\bibfield  {journal} {\bibinfo
   {journal} {International Journal of Mass Spectrometry}\ }\textbf {\bibinfo
  {volume} {362}},\ \bibinfo {pages} {56 } (\bibinfo {year}
  {2014})}\BibitemShut {NoStop}%
\bibitem [{\citenamefont {Gehring}\ \emph {et~al.}(2016)\citenamefont {Gehring}
  \emph {et~al.}}]{Gehring2016221}%
  \BibitemOpen
  \bibfield  {author} {\bibinfo {author} {\bibfnamefont {A.}~\bibnamefont
  {Gehring}} \emph {et~al.},\ }\href {\doibase
  https://doi.org/10.1016/j.nimb.2016.02.012} {\bibfield  {journal} {\bibinfo
  {journal} {Nuclear Instruments and Methods in Physics Research Section B:
  Beam Interactions with Materials and Atoms}\ }\textbf {\bibinfo {volume}
  {376}},\ \bibinfo {pages} {221 } (\bibinfo {year} {2016})},\ \bibinfo {note}
  {proceedings of the \{XVIIth\} International Conference on Electromagnetic
  Isotope Separators and Related Topics (EMIS2015), Grand Rapids, MI, U.S.A.,
  11-15 May 2015}\BibitemShut {NoStop}%
\bibitem [{\citenamefont {Burghardt}(2012)}]{Burghardt2012}%
  \BibitemOpen
  \bibfield  {author} {\bibinfo {author} {\bibfnamefont {T.~P.}\ \bibnamefont
  {Burghardt}},\ }\href {\doibase 10.1117/1.JBO.17.12.126007} {\bibfield
  {journal} {\bibinfo  {journal} {Journal of Biomedical Optics}\ }\textbf
  {\bibinfo {volume} {17}},\ \bibinfo {pages} {126007} (\bibinfo {year}
  {2012})}\BibitemShut {NoStop}%
\bibitem [{\citenamefont {Gell}\ \emph {et~al.}(2006)\citenamefont {Gell},
  \citenamefont {Brockwell},\ and\ \citenamefont {Smith}}]{gell2006handbook}%
  \BibitemOpen
  \bibfield  {author} {\bibinfo {author} {\bibfnamefont {C.}~\bibnamefont
  {Gell}}, \bibinfo {author} {\bibfnamefont {D.}~\bibnamefont {Brockwell}}, \
  and\ \bibinfo {author} {\bibfnamefont {A.}~\bibnamefont {Smith}},\
  }\href@noop {} {\emph {\bibinfo {title} {Handbook of single molecule
  fluorescence spectroscopy}}}\ (\bibinfo  {publisher} {Oxford University Press
  on Demand},\ \bibinfo {year} {2006})\BibitemShut {NoStop}%
\bibitem [{\citenamefont {Stemmer}\ \emph {et~al.}(2008)\citenamefont
  {Stemmer}, \citenamefont {Beck},\ and\ \citenamefont {Fiolka}}]{Stemmer2008}%
  \BibitemOpen
  \bibfield  {author} {\bibinfo {author} {\bibfnamefont {A.}~\bibnamefont
  {Stemmer}}, \bibinfo {author} {\bibfnamefont {M.}~\bibnamefont {Beck}}, \
  and\ \bibinfo {author} {\bibfnamefont {R.}~\bibnamefont {Fiolka}},\ }\href
  {\doibase 10.1007/s00418-008-0506-8} {\bibfield  {journal} {\bibinfo
  {journal} {Histochemistry and Cell Biology}\ }\textbf {\bibinfo {volume}
  {130}},\ \bibinfo {pages} {807} (\bibinfo {year} {2008})}\BibitemShut
  {NoStop}%
\bibitem [{\citenamefont {Fish}(2009)}]{fish2009total}%
  \BibitemOpen
  \bibfield  {author} {\bibinfo {author} {\bibfnamefont {K.~N.}\ \bibnamefont
  {Fish}},\ }\href@noop {} {\bibfield  {journal} {\bibinfo  {journal} {Current
  protocols in cytometry}\ ,\ \bibinfo {pages} {12}} (\bibinfo {year}
  {2009})}\BibitemShut {NoStop}%
\bibitem [{\citenamefont {Habuchi}\ \emph {et~al.}(2005)\citenamefont
  {Habuchi}, \citenamefont {Ando}, \citenamefont {Dedecker}, \citenamefont
  {Verheijen}, \citenamefont {Mizuno}, \citenamefont {Miyawaki},\ and\
  \citenamefont {Hofkens}}]{Habuchi05072005}%
  \BibitemOpen
  \bibfield  {author} {\bibinfo {author} {\bibfnamefont {S.}~\bibnamefont
  {Habuchi}}, \bibinfo {author} {\bibfnamefont {R.}~\bibnamefont {Ando}},
  \bibinfo {author} {\bibfnamefont {P.}~\bibnamefont {Dedecker}}, \bibinfo
  {author} {\bibfnamefont {W.}~\bibnamefont {Verheijen}}, \bibinfo {author}
  {\bibfnamefont {H.}~\bibnamefont {Mizuno}}, \bibinfo {author} {\bibfnamefont
  {A.}~\bibnamefont {Miyawaki}}, \ and\ \bibinfo {author} {\bibfnamefont
  {J.}~\bibnamefont {Hofkens}},\ }\href {\doibase 10.1073/pnas.0500489102}
  {\bibfield  {journal} {\bibinfo  {journal} {Proceedings of the National
  Academy of Sciences of the United States of America}\ }\textbf {\bibinfo
  {volume} {102}},\ \bibinfo {pages} {9511} (\bibinfo {year} {2005})},\ \Eprint
  {http://arxiv.org/abs/http://www.pnas.org/content/102/27/9511.full.pdf}
  {http://www.pnas.org/content/102/27/9511.full.pdf} \BibitemShut {NoStop}%
\bibitem [{\citenamefont {Thomas}\ \emph {et~al.}(2000)\citenamefont {Thomas},
  \citenamefont {Tovey}, \citenamefont {Collins}, \citenamefont {Bootman},
  \citenamefont {Berridge},\ and\ \citenamefont {Lipp}}]{thomas2000comparison}%
  \BibitemOpen
  \bibfield  {author} {\bibinfo {author} {\bibfnamefont {D.}~\bibnamefont
  {Thomas}}, \bibinfo {author} {\bibfnamefont {S.}~\bibnamefont {Tovey}},
  \bibinfo {author} {\bibfnamefont {T.}~\bibnamefont {Collins}}, \bibinfo
  {author} {\bibfnamefont {M.}~\bibnamefont {Bootman}}, \bibinfo {author}
  {\bibfnamefont {M.}~\bibnamefont {Berridge}}, \ and\ \bibinfo {author}
  {\bibfnamefont {P.}~\bibnamefont {Lipp}},\ }\href@noop {} {\bibfield
  {journal} {\bibinfo  {journal} {Cell calcium}\ }\textbf {\bibinfo {volume}
  {28}},\ \bibinfo {pages} {213} (\bibinfo {year} {2000})}\BibitemShut
  {NoStop}%
\bibitem [{\citenamefont {Thompson}\ \emph {et~al.}(2002)\citenamefont
  {Thompson}, \citenamefont {Larson},\ and\ \citenamefont
  {Webb}}]{thompson2002precise}%
  \BibitemOpen
  \bibfield  {author} {\bibinfo {author} {\bibfnamefont {R.~E.}\ \bibnamefont
  {Thompson}}, \bibinfo {author} {\bibfnamefont {D.~R.}\ \bibnamefont
  {Larson}}, \ and\ \bibinfo {author} {\bibfnamefont {W.~W.}\ \bibnamefont
  {Webb}},\ }\href@noop {} {\bibfield  {journal} {\bibinfo  {journal}
  {Biophysical journal}\ }\textbf {\bibinfo {volume} {82}},\ \bibinfo {pages}
  {2775} (\bibinfo {year} {2002})}\BibitemShut {NoStop}%
\end{thebibliography}%

\end{document}